\documentstyle[11pt,astroph,twoside,psfig]{article}

\begin{document}

\title{High Redshift Clusters and Protoclusters}
\author{Marc Postman}
\affil{Space Telescope Science Institute}

\begin{abstract}
Our understanding of the cosmic history of galaxy clusters has
recently been enhanced due to an extensive series of observations
including faint spectroscopic data (especially those obtained at the 
Keck Observatory), deep optical and
NIR imaging from the ground and in space, morphological data from
HST, and new constraints on the evolution of the intracluster medium
from ROSAT and ASCA. 
When such observations are applied to complete, 
objectively derived catalogs of clusters, our constraints on cluster
formation and evolution become quite confined. 
A picture is emerging in which the bulk of cluster formation starts at $z \ge 2$,  
yielding cluster potentials that are well established by $z \sim 1$,
and there is little substantial evolution of the cluster galaxy population since
$z \sim 0.4$. This review talk will summarize the current observational
constraints on the properties and evolution of high redshift clusters
and protoclusters and their implications. 
\end{abstract}

\keywords{clusters of galaxies; galaxy evolution; cluster evolution}

\section{Introduction}

The evolution of galaxy clusters is quite sensitive
to the physical processes which dominate the formation of structure
and to the cosmological parameters. Cluster evolution is inherently
complex both because clusters are not closed systems and because 
the 3 main mass components (dark matter, intracluster gas,
and galaxies) evolve differently. As a consequence,
different cluster parameters evolve on different timescales depending
on the thermal and dissipative properties of the mass component(s)
which most strongly control each cluster parameter. Furthermore,
it is becoming evident that the process of cluster formation extends over
a moderately broad range in redshift.
Given the complexity of the problem, breakthroughs in understanding
the formation and evolution of clusters of galaxies must rely on
observations spread over a large range of wavelength and redshift.

With the advent of the Keck 10m telescopes, the restored resolution of HST 
imagery, improvements in IR arrays, and the enhanced x-ray imaging and 
spectroscopic capabilities of ROSAT and ASCA, constraints on the properties 
of the ICM and the cluster galaxy population have been extended  
out to $z \sim 1$ and beyond. Indeed, cluster candidates
have been identified out to $z \sim 3$ (see Table 1). 
The large temporal baseline these
data cover now allow much tighter constraints on scenarios for cluster 
evolution.

\section{Search Strategies}

The most distant cluster candidates have been found 
by conducting searches in the vicinity of high-z radio galaxies or quasars.
The evidence for the presence of a cluster in these cases is typically based
on fewer than 5 spectroscopically confirmed members plus a statistical
excess of red galaxies or Ly$\alpha$ emitters.

\begin{center}
Table 1. Examples of $z > 1$ Cluster Candidates
\end{center}
\begin{center}
\begin{tabular}{|l|r|l|}
\hline
Name & Redshift & Reference\\
\hline
MRC 0316-257   & 3.14 & Le Fevre et al. 1996\\
QSO 0953+545   & 2.50 & Malkan et al. 1996\\
QSO 1312+4237  & 2.50 & Campos et al. 1999\\
53W002         & 2.39 & Pascarelle et al. 1995\\
QSO 2139-4434  & 2.38 & Francis et al. 1996\\
3C294          & 1.79 & Dickinson et al. 1999\\
RXJ0848+4453$^a$& 1.27 & Stanford et al. 1997; Rosati et al. 1999\\
3C324          & 1.21 & Dickinson 1997\\
AXJ2019+112    & 1.01 & Benitez et al. 1998\\
3C184$^a$      & 1.00 & Deltorn et al. 1997 \\
\hline
\end{tabular}

$^a$ Spectroscopic confirmation based on more than 10 $z$'s
\end{center}

Hall \& Green (1998) have also performed a search around a sample of radio-loud
quasars and have identified 31 possible clusters in the range $1 < z < 2$.
The physical properties of $z > 1$ cluster candidates are not well quantified 
because of the limited amount of spectroscopic data available to date.
While looking for clusters in the vicinities of radio galaxies or quasars is 
fruitful, the resulting samples will naturally suffer from selection effects
associated with limiting ones search to such interesting environments. 
None the less, it appears that overdensities which may be the progenitors
of present day clusters exist at $z \sim 3$ ($\sim 15$\% of the current age
of the universe).

A more complete picture of the properties of high-redshift clusters can
be obtained at $z < 1.3$ through objective searches of wide areas of sky
in optical, NIR, and x-ray passbands.
The advantage of x-ray cluster selection is two-fold: 1) emission from the hot 
intracluster medium (ICM) directly indicates the presence of a gravitationally 
bound system and 2) the ICM comprises 70 to 80\% of the cluster's baryonic mass.
Nearly all x-ray selected high-$z$ clusters are rich and elliptical dominated.
The Extended Medium Sensitivity Survey (EMSS; Gioia et al. 1990a;
Henry et al. 1992) has been used to 
identify clusters out to $z \approx 0.85$ over an area of $\sim 850$ deg$^2$
and the ROSAT Distant Cluster Survey 
(RDCS; Rosati et al.  1998) has been used to find systems out to 
$z \approx 1.3$ over an area of $\sim 30$ deg$^2$.
The RDCS, in particular, includes 100 spectroscopically confirmed clusters. Of 
these, 33\% have $z > 0.4$ and 25\% have $z > 0.5$ (Rosati 1998). 
While past x-ray telescopes have had fairly low effective areas, new 
observatories, like XMM, will provide at least an order of magnitude 
improvement. Equally exciting are developments in the use of the 
Sunyaev-Zeldovich Effect to locate clusters. Mohr et al. (1999) indicate 
that SZE facilities in the near future will be able
to detect $\sim 100\ z > 1$ clusters per year.

Searching for distant clusters in the optical and NIR, however, also has
significant advantages. From a practical point of view, there are more 
telescopes and larger area mosaic cameras available in the optical/NIR than in
x-rays. Optical/NIR searches will also find clusters spanning a wider range
of x-ray luminosity and total mass ({\it e.g.}, Holden et al. 1997). 
Although the spurious detection 
rate at high-$z$ can be $\sim 30$\%, the use of photometric 
redshifts can dramatically reduce the number of false positives. 
Some of the largest area and deepest optical/NIR distant cluster surveys
include the Palomar Distant Cluster Survey (5.1 deg$^2$; Postman et al. 
1996), the ESO Imaging Survey (12 deg$^2$; Scodeggio et al. 1999)
The Deeprange Survey
(16 deg$^2$; Postman et al. 1998), and the NOAO Deep-Wide Survey (18 deg$^2$;
Jannuzi \& Dey 1999).

An optimal strategy, of course, is to combine x-ray and optical/NIR 
data obtained over the same region of sky.  
This allows a full assessment of the selection biases to be made
and is likely to reveal subtle effects which can be important in
interpreting, for example, the abundance of high-$z$ clusters.
The benefits of such joint searches are already
being realized: Scharf et al. (1999) used 
22 deep ROSAT PSPC fields as targets for deep optical imaging to 
study the effects of optical and x-ray selection on derived cluster evolution
and to look for correlations in the large-scale distribution of diffuse x-ray
emission and the galaxy distribution.  
Preliminary results include the possible first
direct detection of x-ray emission from an intercluster
filament at $z \sim 0.4 - 0.5$. 
Stanford et al (1997) and Rosati et al (1999) 
have identified a supercluster at $z = 1.27$ in the Lynx field
which was initially detected
in the NIR (K-band) and, subsequently, in x-rays.

\section{Cluster Abundance}

The abundance of clusters as a function of redshift is one of the fundamental
constraints on both structure formation and cosmological models. The 
space density of clusters at $z > 0.5$, for example, 
is highly sensitive to $\Omega_m$ (Bahcall, Fan, \& Renyue 1997; 
Donahue \& Voit 1999). 
Present observational constraints from x-ray surveys (followed up by
optical spectroscopy) indicate that the 
comoving space density of clusters 
per unit $L_x$ is invariant out to at least $z = 0.8$ for
systems with $L_x \le 3 \times 10^{44}$ erg sec$^{-1}$ 
(Henry et al. 1992; Rosati et al. 1998). For more luminous (massive) clusters, 
mild negative evolution has been reported (Gioia et al. 1990b; 
Henry et al. 1992; Vikhlinin et al. 1998) although the deficit,
expressed in absolute numbers, is only a dozen or so EMSS clusters 
at $z > 0.4$ (small enough that one might worry about subtle selection biases
at low x-ray surface brightness levels
in the existing surveys). The distribution of poor to moderately rich 
optically selected clusters is also consistent with
a constant comoving space density to at least $z = 0.6$ 
(Postman et al.  1996; Holden et al. 1998). 

At $z > 1$, our constraints on cluster abundances 
presently suffer from a lack of data. There are at 
least 5 known clusters with $0.75 < z < 1.3$ that have velocity dispersions 
in the range $700 \le \sigma_{1D} \le 1400$ km sec$^{-1}$. At least two of 
these, MS-1137 ($z=0.78$) and MS-1054 ($z=0.83$), have relatively
high kinetic gas temperatures ($T_x$) -- 5.7 keV and 12.4 keV, respectively 
(Donahue et al. 1998, 1999). The existence of massive 
($> 5 \times 10^{14}$ M$_{\odot}$)
clusters at $z \sim 1$ is, thus, no longer in doubt. Interestingly, 
the space density of $z \sim 1$ clusters inferred from
the RDCS coverage of the Lynx region is $\sim 2.4 \times 10^{-6}
h^3$ Mpc$^{-3}$ (Rosati 1999), within a factor of 2 of the density $z \sim 3$ 
structures delineated by Lyman break galaxies (Steidel et al. 1998). Whether this 
is indicative of an evolutionary connection or mere coincidence remains to 
be decided.  
\begin{figure*}
\centering\mbox{\psfig{figure=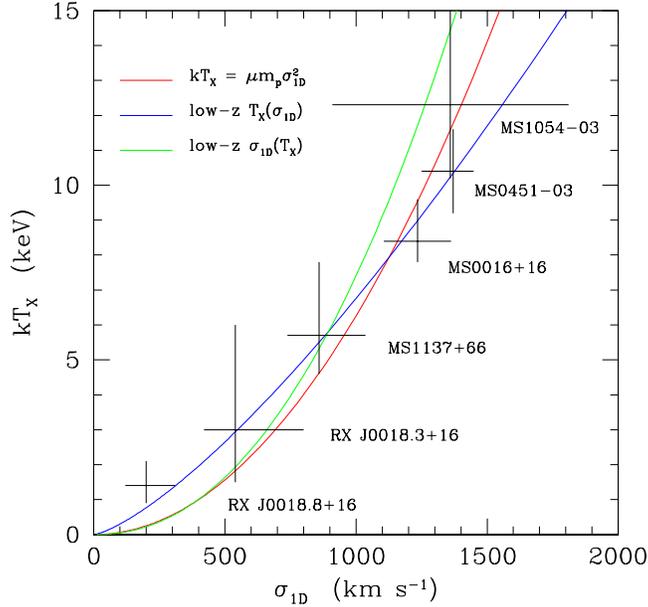,height=4.5in}}
\caption[]{The Donahue et al. (1998)
$T_x$ -- $\sigma_{1D}$ relationship for 6 rich clusters in the
range $0.5 < z < 0.83$. The best fit low-$z$ relationships are shown for
comparison. The distant cluster data are consistent with that seen at
$z < 0.2$.}
\label{Txsig}
\end{figure*}

\section{Evolution of the Gravitational Potentials of Clusters}

The evolution of the ICM and its correlation with global
cluster kinematics provide direct constraints on the growth of
the gravitational potentials which we call clusters. The relationship
between $L_x$ and $T_x$ in low-$z$ clusters is 
remarkably tight but is somewhat steeper than that predicted
by bremsstrahlung emission from a population of virialized,
structurally identical clusters with constant gas fraction 
(Arnaud \& Evrard 1999). However, with reasonable constraints on
cluster structure, the same authors find the fractional variation in
cluster gas fraction is $<15$\%.
Mushotsky \& Scharf (1997) demonstrate that the $L_x - T_x$ 
relationship exhibits
no significant evolution out to $z \approx 0.4$. Donahue et al. (1999) have
extended this work out to $z \sim 0.9$: if the evolution of the relation is
parameterized as $L_x = T_x^{\alpha}(1+z)^A$, then they find that
$A \ge 1.5$ is rejected with greater than $3\sigma$ confidence 
(for $q_o = 0.5$). Values of $A = 2 - 3$ would be required to explain
the lack of evolution in the x-ray luminosity function cited above if
$\Omega_m = 1$. Donahue et al. (1998) have further shown 
(see Figure~\ref{Txsig}) that the
relation between the cluster velocity dispersion ($\sigma_{1D}$) and
$T_x$ is invariant out to $z \sim 0.8$. Cluster potentials are clearly
well established in the universe by $z \sim 0.9$ and, on average, the
x-ray properties of the ICM are similar to those in current epoch clusters.
 
The distribution of the optically luminous mass in clusters, as delineated
by the member galaxies, may be experiencing more recent evolution than the ICM.
Clusters exhibit significantly more asymmetry in their galaxy
distribution at $z > 0.7$ than at the present epoch (Lubin \& Postman 1996) --
the observed profiles are inconsistent with azimuthal symmetry
at the 99.9\% confidence level, in strong contrast with the situation
at $z < 0.3$. In some cases, like
MS-1054, the clumpiness seen in the galaxy distribution is 
also seen in the x-ray brightness distribution (Donahue et al. 1998) and
in a mass map based upon weak lensing distortions (Hoekstra, Franx, \&
Kuijken 1999) --
characteristic of recent merger activity. Indeed, mergers of group-size
clumps at $z \sim 1$ may be the origin of some of
the current epoch richness class 1 and 0 clusters (Lubin et al. 1998;
Gioia et al. 1999).
However, the majority of the known high-$z$ clusters appear to have
been in existence since at least $z \sim 2$, as observations discussed below
suggest.

\section{Constraining the Epoch of Cluster Formation}

The processes which control the formation of clusters leave observable 
signatures in the evolution of the morphological and spectrophotometric 
properties of cluster galaxies. This is a key reason why
observational work in this area has been a major component of
recent extragalactic telescope programs. 
The evolution of the mass function of cluster galaxies, in particular,
provides critical constraints on and tests of cluster formation scenarios.
At high-$z$, cluster galaxy mass determinations are difficult 
to obtain but the K-band cluster galaxy luminosity function (KLF) can provide
a reliable substitute because it probes the total stellar mass component
and is not strongly sensitive to the instantaneous star formation rate
(e.g., see Gavazzi, Pierini, Boselli 1996). De Propris et al. (1999)
have derived the KLF for 38 clusters at $0.1 < z < 1$. Their main
result is that the KLF departs from no-evolution predictions at $z > 0.4$
however the changes observed are consistent with simple, passive evolution
(aging of the existing stellar population)
and a narrow formation epoch around $z = 2$ (if $\Lambda = 0.7$)  
or $z = 3$ (if $\Lambda = 0$).

Comparison of the broadband colors and spectral features of the early
type cluster members at $0.76 \le z \le 0.92$ with spectral synthesis
models suggests these galaxies are old (mean ages $\sim3\pm2$ Gyr, at the
observed redshifts) implying a relatively early formation at $z > 2$ as well 
(e.g., Bower, Kodama, Terlevich 1998; 
Postman, Lubin, Oke 1998; Stanford, Eisenhardt, Dickinson 1998). 
Such observations also suggest that the mass-to-light 
ratios of the early type cluster
galaxies have evolved passively since at least $z \sim 1.2$ 
(Kelson et al. 1997; van Dokkum et al. 1998). Taken in concert with
the results on the KLF evolution, one may conclude that the mass
function of cluster galaxies has remained roughly invariant since $z \sim 1.2$.
 
An additional constraint of the duration of the cluster galaxy
formation era comes from the optical/NIR color-magnitude relations for
the red galaxy population which are well-defined and exhibit remarkably low
scatter in clusters from $0 < z < 1$ (Stanford, Eisenhardt, Dickinson 1998). 
This places
a stringent constraint on their formation synchronicity of $\Delta t \le 4$
Gyr (roughly the time between $z = 10$ and $z = 1.5$ in a 
$\Omega_o = 0.2,\ h = 0.6,\ \Lambda = 0$ cosmology). 
The coeval nature of cluster elliptical evolution is also reflected in 
observations which are consistent with exponentially decaying 
star formation rates with relatively short
$e$-folding times ($0.1 < \tau < 0.6$ Gyr; Postman, Lubin, Oke 1998).

There is some evidence (Fuller, West, Bridges 1999) that the brightest
cluster galaxies are preferentially aligned with the global cluster
galaxy distribution, an effect also suggestive of an early formation epoch.

The metallicities of cluster ellipticals in the redshift range $0.5 < z < 1$
are consistent, on average, with the close to solar values observed
in current epoch ellipticals.
Similarly, there does not appear to have been much change in the
metallicity of the ICM ($\sim 0.2 - 0.45$ solar)
between now and $z \sim 0.8$ (Donahue et al. 1999).
This suggests reprocessing of the baryonic mass component of clusters
had been on-going for a few Gyr prior to current lookback time.
Hierarchical models do predict, however, that some star formation 
activity should be occurring in clusters since $z \sim 1$. This is indeed
seen in the spiral members and in the surrounding field galaxy population.
For example, the fraction of cluster members with strong OII emission 
(EW $>$ 15\AA), a reliable star formation indicator, 
increases by a factor of 3 - 4 between now and $z \sim 0.92$
(Balogh et al. 1998; Postman, Lubin, Oke 1998). 
The percentage of cluster galaxies
with post-starburst spectral features increases nearly tenfold between
now and $0.3 < z < 0.5$ (Dressler et al. 1999).

\section{Evolution of the Morphological Mix}

\begin{figure*}
\centering\mbox{\psfig{figure=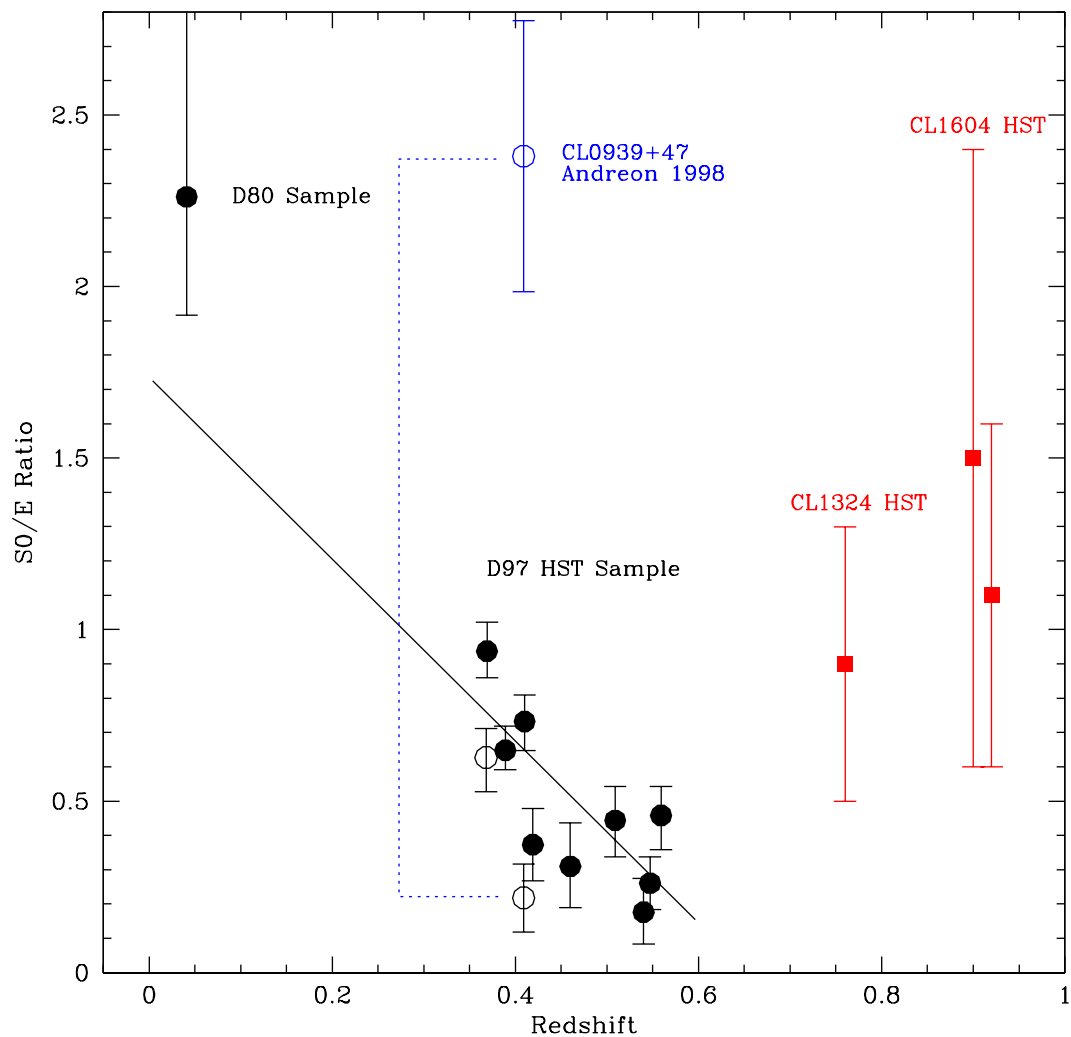,height=6.0in}}
\caption[]{The evolution of the S0/E ratio based on WFPC and WFPC2
imaging of clusters in the range $0.3 < z < 0.95$. The low-$z$ reference
is from Dressler (1980). The intermediate redshift results are from
Dressler et al. 1997. Andreon (1998) re-analyzed their data 
and finds higher S0/E ratios which are consistent with the local
value -- his ratio for CL0939+47 is highlighted. 
The three high-$z$ clusters
are at $z=0.76,\ 0.90,\ {\rm and}\ 0.92$ (Lubin et al. 1999).}
\label{morfz}
\end{figure*}

Using the high angular resolution imaging provided by
the Hubble Space Telescope, several teams 
(e.g., Dressler et al. 1997; Oemler et al. 1997; Smail et al. 1997;
Lubin et al. 1999) have conducted morphological
surveys of clusters in from $0.3 < z < 1$. Upon comparison with
similar ground-based studies of low-$z$ clusters, it appears that 
the distribution of cluster galaxy morphologies has undergone rather
substantial evolution between $0.4 < z < 1$ but remains relatively
invariant between $z = 0.4$ and the present epoch. 
One indicator that has been used to gauge the evolution is the
ratio of the number of lenticulars (S0) to ellipticals in the central regions
of clusters. Figure~\ref{morfz} summarizes the current constraints on
the redshift dependence of the S0/E ratio in clusters observed by
Dressler et al. 1997 (D97) and Lubin et al. 1999. 
Although the D97 results have been widely cited as evidence for 
substantial morphological evolution even since $z = 0.4$, re-analysis
of the data by Andreon (1998) and new data at higher $z$ from
Lubin et al. 1999 suggest that the metamorphosis may not be as dramatic
as originally thought. However, the high-$z$ data from Lubin et al. also
find that the while ellipticals still preferentially reside in 
the highest density environments, the Spiral/S0 
morphology-density relation at $z > 0.5$ is much less well-defined
than it is now. 

There is further evidence that morphological
modifications are occurring at least as recently as $z \sim 0.3 - 0.4$:
Dressler et al. (1999) find that the OII equivalent widths, for a given
morphological type, are lower in $0.3 < z < 0.5$ 
cluster galaxies than in the field and that the actively star-forming galaxies
in these clusters have a more extended spatial distribution than the
non-active galaxies.  The kinematic properties of the early and
late type cluster galaxies appear to differ as well, although these
differences exist both in the present epoch and at $z \sim 1$
(Adami et al. 1998; Lubin et al. 1999). Specifically, the spiral population
tends to have a higher velocity dispersion than the elliptical members
suggestive of an on-going spiral infall process.
Poggianti et al. (1999) find, however, evidence for widespread cessation
of star formation activity in intermediate $z$
clusters over a relatively short ($\sim 1$ Gyr) timescale. Specifically,
90\% of the spiral cluster members they studied show spectral signatures of
either enhanced or suppressed star formation relative to local spirals.

\section{A Possible Scenario}

I will now propose a possible evolutionary sequence  
which incorporates the myriad of constraints derived
from observations of clusters. Note that some of the steps below have
not been observationally confirmed or are still controversial!
\begin{itemize}
\item In a universe with $\Omega_m < 1$ (and possibly $\Lambda \ne 0$),
protoclusters form at $z > 3$.
\item The sites of formation are located at the intersections of filamentary
matter flows and the first cluster galaxies form during
the first generation of matter crossings
({\it This is a conjecture based solely on N-body simulations and
popular dark matter models}).
\item The richest, current epoch clusters formed first. Some of the poorer
clusters seen today may have developed via group-group mergers since
$z \sim 1$.
\item Primordial ICM shocks and begins to emit x-rays at $z \sim 2$
(and perhaps earlier). Enrichment of the ICM most likely occurs in
the $z > 1$ era. From $z \sim 1$ to now, there is little
evolution of the ICM.
\item The brightest cluster galaxies grow via cannibalism until
$z \sim 1.5$. Most merger activity ceases by $z \sim 1$ and subsequent
evolution is passive. Other massive ellipticals assemble prior to
$z \sim 2$ and are the first to reach dynamical equilibrium with the
cluster potential.
\item The most active periods of star formation within the cluster
occur at $z > 1$.  Most star formation is quenched, however, by 
$z \sim 0.4 - 0.5$.
\item Infall of spirals results in morphological and color gradients within
the cluster. This process continues up to the present epoch.
\item The S0 and dwarf elliptical populations develop within the cluster
core, certainly by $z \sim 0.5$ and more likely by $z \sim 1$. The likely
relevant processes involved are ram pressure stripping, mergers, and
tidal stress. S0's may descend from high surface brightness spirals,
dE's from low surface brightness spirals (Moore at al. 1998).
\end{itemize}
There are notable exceptions to the above scenario such as
low-$z$ spiral rich clusters (e.g., Virgo, Hercules) and low-$z$
irregular clusters (e.g., Abell 1185), which are probably still 
dynamically young, suggesting that some cluster evolution is still occurring
at the present epoch. Furthermore, our knowledge of cluster evolution
at $z > 1$ is still quite rudimentary. Thus, while great strides have been made,
there remain many steps to go before our understanding of the cluster 
formation process is complete. Some of the observational programs
which will take us farther towards this goal 
are now, or soon will be, underway. These include more complete
and larger $z > 1$ cluster samples, more objective and precise studies of
the $z = 0$ cluster population (e.g., the Sloan Digital Sky Survey, the
2dF survey, the REFLEX survey), improved x-ray observations from XMM and
Chandra and optical/IR observations with HST (ACS, WF3) and SIRTF,
extended spectroscopic studies of high-$z$ clusters using the growing
suite of 8 -- 10m ground-based telescopes, construction of mass-selected
cluster catalogs from SZE surveys, and ultra deep
21 cm searches for protoclusters at $z > 2$ using the Giant Meter-wave Radio
Telescope.

\acknowledgments
I thank the SOC for their generous travel support which made attendance
of this meeting possible.
I also wish to thank Megan Donahue and Caleb Scharf for providing
results based on their x-ray observations of clusters in advance of publication.


\begin{references}
\reference{Adami, C. et al. 1998, A\&A, 331, 439}
\reference{Andreon, S. 1998, ApJ, 501, 533}
\reference{Arnaud, M., Evrard, A. 1999, MNRAS, 305, 631}
\reference{Bahcall, N., Fan, X., Renyue, C. 1997, ApJ, 485, L53}
\reference{Balogh, M. et al. 1998, ApJ, 504, L75}
\reference{Benitez, N. et al. 1998, astro-ph/9812218} 
\reference{Bower, R., Kodama, T., Terlevich, A. 1998, MNRAS, 299, 1193}
\reference{Campos, A. et al., 1999, ApJ, 511, L1} 
\reference{De Propris, R. et al. 1999, AJ, 118, 719} 
\reference{Deltorn, J. M. et al. 1997, ApJ, 483, L21} 
\reference{Dickinson, M. 1997, in {\it HST and the High Redshift Universe,}
eds.\ N.\ Tanvir, A.\ Aragon-Salamanca, and J.V.\ Wall, 
(Singapore: World Scientific), p. 207}
\reference{Dickinson, M. et al. 1999, ApJ, submitted}
\reference{Donahue, M. \& Voit, M. 1999, ApJ, 523, L137}
\reference{Donahue, M. et al. 1998, ApJ, 502, 550}
\reference{Donahue, M. et al. 1999, ApJ, in press, also astro-ph/9906295}
\reference{Dressler, A. 1980, ApJ, 236, 351}
\reference{Dressler, A. et al. 1997, ApJ, 490, 577}
\reference{Dressler, A. et al. 1999, ApJS, 122, 51}
\reference{Francis, P. et al. 1996, ApJ, 457, 490} 
\reference{Fuller, T., West, M., Bridges, T. 1999, ApJ, 519, 22}
\reference{Gavazzi, G., Pierini, D., Boselli, A. 1996, A\&A, 312, 397}
\reference{Gioia, I. et al. 1990a, ApJS, 116, 247} 
\reference{Gioia, I. et al. 1990b, ApJ, 365, 35} 
\reference{Gioia, I. et al. 1999, AJ, 117, 2608} 
\reference{Hall, P., Green, R. 1998, ApJ, 507, 558} 
\reference{Henry, J. et al. 1992, ApJ, 386, 408} 
\reference{Hoekstra, H., Franx, M., Kuijken, K. 1999, ApJ, in press}
\reference{Holden, B. et al. 1997, AJ, 114, 1701} 
\reference{Holden, B. et al. 1998, AAS, 193, 3817} 
\reference{Jannuzi, B., Dey, A. 1999, AAS, 194, 8803} 
\reference{Kelson, D. et al. 1997, ApJ, 478, L13}
\reference{Le Fevre, O. et al. 1996, ApJ, 471, L11} 
\reference{Lubin, L., Postman, M. 1996, AJ, 111, 1795}
\reference{Lubin, L. et al. 1998, AJ, 116, 584} 
\reference{Lubin, L. et al. 1999, AJ, submitted} 
\reference{Malkan, M., Teplitz, H., Mclean, I. 1996, ApJ, 468, L9} 
\reference{Mohr, J. et al. 1999, astro-ph/9905256} 
\reference{Oemler, A., Dressler, A., Butcher, H. 1997, ApJ, 474, 561}
\reference{Pascarelle, S.M. et al. 1996, ApJ, 456, L21} 
\reference{Poggianti, B. et al. 1999, ApJ, 518, 576}
\reference{Postman, M. et al. 1996, AJ, 111, 615} 
\reference{Postman, M. et al. 1998, ApJ, 506, 33} 
\reference{Postman, M., Lubin, L., Oke, J. B. 1998, AJ, 116, 560} 
\reference{Rosati, P. 1998, in {\it Wide Field Surveys in Cosmology},
eds. S. Colombi, Y. Mellier, B. Raban, p. 219} 
\reference{Rosati, P. et al. 1998, ApJ, 492, L21} 
\reference{Rosati, P. 1999, private comm.}
\reference{Rosati, P. et al. 1999, AJ, 118, 76} 
\reference{Scharf, C. et al. 1999, ApJ, in press}
\reference{Scodeggio, M. et al. 1999, A\&AS, 137, 83} 
\reference{Smail, I. et al. 1997, ApJS, 110, 213}
\reference{Stanford, S. et al. 1997, AJ, 114, 2232} 
\reference{Stanford, S., Eisenhardt, P., Dickinson, M. 1998, Apj, 492, 461}
\reference{Steidel, C. et al. 1998, ApJ, 492, 428}
\reference{van Dokkum, P. et al. 1998, ApJ, 504, L17}
\reference{Vikhlinin, A. et al. 1998, ApJ, 502, 558}
\end{references}
\end{document}